\documentclass[12pt]{article}
\usepackage{amsmath}
\begin{document}
\thispagestyle{empty}
\vspace*{5cm}
\begin{center}
\textbf{\LARGE Weyl equation for temperature fields \\
induced by attosecond laser pulses}
\vspace{2cm}

{\Large Janina Marciak-Kozlowska, Miroslaw Kozlowski}

\vspace{2cm}
Institute of Electron Technology, Al. Lotnik\'{o}w 32/46, 02-668 
Warszawa, Poland
\end{center}
\newpage
\begin{abstract}
In this paper the Weyl equation for temperature field induced by laser beam interaction with matter is proposed and solved. Depending on the scattering mechanism the temperature field oscillate or is damped.\\
\textbf{Key words}: Thermal process, Weyl equation.
fields.
\end{abstract}
\newpage
\section{Derivation of the 1+1 dimensional Dirac and Weyl equations for thermal 
processes}
As pointed in papers{\nobreakspace}\cite{1,2} spin-flip occurs only when there 
is more than one dimension in space. Repeating the discussion 
of deriving the Dirac equation{\nobreakspace}\cite{3} for the case of one 
spatial dimension, one easily finds that the Dirac matrices {$\alpha$} and {${\beta}$} 
 are reduced to $2\times2$ matrices that can be represented by the Pauli 
matrices{\nobreakspace}\cite{3}. This fact simply implies that if there is 
only one spatial dimension, there is no spin. It should be instructive 
to show explicitly how to derive the 1+1 dimensional Dirac equation.\\
As discussed in textbooks{\nobreakspace}\cite{2,3},  a wave equation 
that satisfies relativistic covariance in space-time as well 
as the probabilistic interpretation should have the form:
\begin{equation}
i\hbar \frac{\partial }{\partial t} \Psi \left( x,t\right) =\left[
c{{\alpha}}\left( -i\hbar \frac{\partial }{\partial x} \right) +{\beta}m_{0} c^{2} \right]
\Psi \left( x,t\right).\label{eq1} 
\end{equation}
To obtain the relativistic energy-momentum relation 
$E^{2} =\left( pc\right) ^{2} +m_{0}^{2} c^{4} $
 we postulate that{\nobreakspace}(\ref{eq1}) coincides with the Klein-Gordon 
equation
\begin{equation}
\left[ \frac{\partial ^{2} }{\partial \left( ct\right) ^{2} }
-\frac{\partial ^{2} }{\partial x^{2} } +\left( \frac{m_{0} c}{\hbar }
\right) ^{2} \right] \Psi \left( x,t\right) =0.\label{eq2}
\end{equation}
By comparing (\ref{eq1}) and (\ref{eq2}) it is easily seen that $\alpha$ 
 and $\beta$
 must satisfy
\begin{equation}
{\alpha}^{2} -{\beta}^{2} =1,\quad \quad {\alpha}{\beta}+{\beta}{\alpha}=0.\label{eq3}
\end{equation}
Any two of the Pauli matrices can satisfy these relations. Therefore, 
we may choose 
${\alpha}={\sigma}_{x} $
 and 
${\beta}={\sigma}_{z} $
 and we obtain:
\begin{equation}
i\hbar \frac{\partial }{\partial t} \Psi \left( x,t\right) =\left[ c\sigma_{x}
\left( -i\hbar \frac{\partial }{\partial x} \right) +{\sigma}_{z} m_{0} c^{2}
\right] \Psi \left( x,t\right),\label{eq4}
\end{equation}
where 
$\Psi (x,t)$
 is a 2-component spinor.
 
The Eq.{\nobreakspace}(\ref{eq4}) is the Weyl  equation. 
We perform a phase transformation on 
$\Psi \left( x,t\right) $
 letting 
$u\left( x,t\right) =\exp \left( \frac{imc^{2} t}{\hbar } \right) \Psi
\left( x,t\right). $
Call $u$'s upper (respectively, lower) component 
$u_{+} \left( x,t\right) $, 
$u_{-} \left( x,t\right) $; it follows from{\nobreakspace}(\ref{eq4}) that 
$u_{\pm } $
 satisfies
\begin{equation}
\frac{\partial u_{\pm } \left( x,t\right) }{\partial t} =\mp
c\frac{\partial u_{\pm } }{\partial x} +\frac{im_{0} c^{2} }{\hbar }
\left( u_{\pm } -u_{\mp } \right).\label{eq5}
\end{equation}
Following the physical interpretation of the equation{\nobreakspace}(\ref{eq5}) 
it describes the relativistic particle (mass{\nobreakspace}$m_{0} $) propagates at the speed of light $c$ and with a certain \textit{chirality}
(like a two component neutrino) except that at random times it 
flips both direction of propagation (by 180$^{0}$) and chirality.

In monograph{\nobreakspace}\cite{4} we considered a particle moving on the 
line with fixed speed $w$ and supposed that from time to time 
it suffers a complete reversal of direction, 
$u\left( x,t\right) \Leftrightarrow v\left( x,t\right) $, where 
$u\left( x,t\right) $
 denotes the expected density of particles at $x$ and at time $t$ 
moving to the right, and 
$v\left( x,t\right) \equiv $
 expected density of particles at$x$ and at time $t$ moving 
to the left. In the following we perform the change of the abbreviation
\begin{eqnarray}
u\left( x,t\right) &\rightarrow &u_{+} ,\label{eq6}\\ 
v\left( x,t\right) &\rightarrow &u_{-}.\nonumber
\end{eqnarray}
Following the results of the paper{\nobreakspace}\cite{4} we obtain for the 
$u_{\pm } \left( x,t\right) $
 the following equations\\
\begin{eqnarray}
\frac{\partial u_{+} }{\partial t} &=&-w\frac{\partial u_{+} }{\partial
x} -\frac{w}{\lambda } \left( \left( 1-k\right) u_{+} -ku_{-} \right) ,\label{eq7}\\ 
\frac{\partial u_{-} }{\partial t} &=&w\frac{\partial u_{-} }{\partial
x} +\frac{w}{\lambda } \left( ku_{+} +\left( k-1\right) u_{-} \right).\nonumber
\end{eqnarray}
In equation{\nobreakspace}(\ref{eq7}) $k_x$ denotes the number of the 
particles which are moving in left (right) direction after the 
scattering at $x$. The mean free path for scattering is equal 
$\lambda $, 
$\lambda =w\tau $, where $\tau $
 is the relaxation time for scattering.
 
Comparing formulae{\nobreakspace}(\ref{eq5}) and (\ref{eq7}) we conclude that the shapes 
of both equations are the same. In the subsequent we will call 
the set of the equations{\nobreakspace}(\ref{eq7}) \textit{the Weyl equation} for 
the particles with velocity $w$, mean free path~$\lambda $.
For thermal processes we define 
$T_{+,-} \equiv $
 the temperature of the particles with chiralities + and{\nobreakspace}-- 
respectively and with analogy to equation{\nobreakspace}(\ref{eq7}) we obtain:
\begin{eqnarray}
\frac{\partial T_{+} }{\partial t} &=&-w\frac{\partial T_{+} }{\partial
x} -\frac{w}{\lambda } \left( \left( 1-k\right) T_{+} -kT_{-} \right) ,\label{eq8}\\ 
\frac{\partial T_{-} }{\partial t} &=&w\frac{\partial T_{-} }{\partial
x} +\frac{w}{\lambda } \left( kT_{+} +\left( k-1\right) T_{-} \right) ,\nonumber
\end{eqnarray}
where 
$\frac{w}{\lambda } =\frac{1}{\tau } $.

In one dimensional case we introduce one dimensional cross section for 
scattering
\begin{equation}
\sigma \left( x,t\right) =\frac{1}{\lambda \left( x,t\right) }\label{eq9}
\end{equation}

\section{The solution of the Weyl equation for stationary temperatures 
in one dimensional wire}
In the stationary state thermal transport phenomena 
$\frac{\partial T_{+,-} }{\partial t} =0$
 and Eq.{\nobreakspace}(\ref{eq8}) can be written as
\begin{eqnarray}
\frac{dT_{+} }{dx} &=&-\sigma \left( \left( 1-k\right) T_{+} +kT_{-}
\right) , \label{eq10}\\ 
\frac{dT_{-} }{dx} &=&\sigma \left( k-1\right) T_{-} +\sigma kT_{+} .\nonumber
\end{eqnarray}
After the differentiation of the equation{\nobreakspace}(\ref{eq9}) we obtain 
for 
$T_{+} (x)$
\begin{eqnarray*}
\frac{d^{2} T_{+} }{dx} &-&\frac{1}{\sigma k} \frac{d}{dx} \left( \sigma
k\right) \frac{dT_{+} }{dx} +\\
&&T_{+} \left[ \sigma ^{2} \left( 2k-1\right)
+\frac{d\sigma }{dx} \left( 1-k\right) +\frac{\sigma \left( k-1\right)
}{\sigma k} \frac{d\left( \sigma k\right) }{dx} \right] =0.
\end{eqnarray*}
Equation{\nobreakspace}(\ref{eq1}0) can be written in a compact form
$$
\frac{d^{2} T_{+} }{dx^{2} } +f\left( x\right) \frac{dT_{+} }{dx}
+g\left( x\right) T_{+} =0,
$$
where
\begin{eqnarray}
f\left( x\right) &=&-\frac{1}{\sigma } \left( \frac{\sigma }{k}
\frac{dk}{dx} +\frac{d\sigma }{dx} \right) ,\label{eq11}\\ 
g\left( x\right) &=&\sigma ^{2} \left( x\right) \left( 2k-1\right)
-\frac{\sigma }{k} \frac{dk}{dx} .\nonumber
\end{eqnarray}
In the case for constant 
$\frac{dk}{dx} =0$
we obtain
\begin{eqnarray}
f\left( x\right) &=&-\frac{1}{\sigma } \frac{d\sigma }{dx} ,\label{eq12}\\ 
g\left( x\right) &=&\sigma ^{2} \left( x\right) \left( 2k-1\right) .\nonumber
\end{eqnarray}
With functions $f(x)$, $g(x)$ described by formula{\nobreakspace}(\ref{eq12}) 
the general solution of Eq.{\nobreakspace}(\ref{eq1}2) has the form:
\begin{equation}
T_{+} \left( x\right) =C_{1} e^{\left( 1-2k\right) ^{\frac{1}{2}  } \int
\sigma \left( x\right)dx} +C_{2} e^{-\left( 1-2k\right) ^{\frac{1}{2}  }
\int \sigma \left( x\right) dx}\label{eq13}
\end{equation}
and
\begin{eqnarray}
T_{-} (x) &=&\frac{\left[ \left( 1-k\right) +\left(
1-2k\right) ^{\frac{1}{2} } \right] }{k} \times \label{eq14}\\ 
&&={}\left[ C_{1} e^{\left( 1-2k\right) ^{\frac{1}{2} } \int \sigma
\left( x\right) dx}+\frac{\left( 1-k\right) -\left( 1-2k\right)
^{\frac{1}{2} } }{\left( 1-k\right) +\left( 1-2k\right) ^{\frac{1}{2} } }
C_{2} e^{-\left( 1-2k\right) ^{\frac{1}{2} } \int \sigma \left( x\right)
dx }  \right] .\nonumber
\end{eqnarray}
The formulae (\ref{eq13}) and (\ref{eq14}) describe three different mode for heat 
transport. For 
$k=\frac{1}{2} $
 we obtain 
$T_{+} \left( x\right) =T_{-} \left( x\right) $
 while for 
$k>\frac{1}{2} $, i.e. for heat carrier generation 
$T_{+} \left( x\right) $
 and 
$T_{-} \left( x\right) $
 oscillate for 
$\left( 1-2k\right) ^{\frac{1}{2} } $
 is a complex number. For 
$k<\frac{1}{2} $
i.e. for absorption 
$T_{+} \left( x\right) $ and 
$T_{-} \left( x\right) $
decrease as the function of $x$.\\
In the subsequent we will consider the solution of Eq.{\nobreakspace}(\ref{eq9}) 
for Cauchy conditions:
\begin{equation}
T_{+}(0) =T_0 , \qquad  T_{-}( a) =0.\label{eq15}
\end{equation}
Boundary conditions{\nobreakspace}(\ref{eq15}) describes the generation of heat 
carriers by illuminating the left end of one dimensional slab 
(with length \textit{a}) by laser pulse.
From formulae (\ref{eq13}) and (\ref{eq14}) we obtain:
\begin{eqnarray}
T_{+} (x) &=&\frac{2T_0 e^{[f(0) -f(a)]}}{1+\beta e^{2[f(0) -f(a)] } } 
\times \label{eq16}\\ 
&&\mbox{}\frac{\left( 1-2k\right) ^{\frac{1}{2} } \cosh \left[
f\left( x\right) -f\left( a\right) \right] +\left( k-1\right) \sinh \left[
f\left( x\right) -f\left( a\right) \right] }{\left( 1-2k\right)
^{\frac{1}{2} } -\left( k-1\right) },\nonumber
\end{eqnarray}
\begin{equation}
T_{-} \left( x\right) =\frac{2T_{0} e^{2\left[ f\left( 0\right) -f\left(
a\right) \right] } \left[ \left( k-1\right) +\left( 1-2k\right)
^{\frac{1}{2} } \right] \sinh \left[ f\left( x\right) -f\left( a\right)
\right] }{\left( 1+\beta e^{-2\left[ f\left( a\right) -f\left( 0\right)
\right] } \right) k}.\label{eq17}
\end{equation}
In formulae{\nobreakspace}(\ref{eq16}) and (\ref{eq17})
\begin{equation}
\beta =\frac{\left( 1-2k\right) ^{\frac{1}{2} } +\left( k-1\right)
}{\left( 1-2k\right) ^{\frac{1}{2} } -\left( k-1\right) }\label{eq18}
\end{equation}
and
\begin{eqnarray}
f\left( x\right) &=&\left( 1-2k\right) ^{\frac{1}{2} } \int \sigma
\left( x\right) dx ,\\ \nonumber
f\left( 0\right) &=&\left( 1-2k\right) ^{\frac{1}{2} } \left[ \int
\sigma \left( x\right) dx \right] _{0} ,\\ \label{eq19}
f\left( a\right) &=&\left( 1-2k\right) ^{\frac{1}{2} } \left[ \int
\sigma \left( x\right) dx \right] _{a} .\nonumber
\end{eqnarray}
With formulae (\ref{eq16}) and (\ref{eq17}) for 
$T_{+} \left( x\right) $
and 
$T_{-} \left( x\right) $
we define the asymmetry $A(x)$ of the temperature $T(x)$
\begin{equation}
A\left( x\right) =\frac{T_{+} \left( x\right) -T_{-} \left( x\right)
}{T_{+} \left( x\right) +T_{-} \left( x\right) },\label{eq20}
\end{equation}
\begin{equation}
A\left( x\right) =\frac{\frac{\left( 1-2k\right) ^{\frac{1}{2} } }{\left(
1-2k\right) ^{\frac{1}{2} } -\left( k-1\right) } \cosh \left[ f\left(
x\right) -f\left( a\right) \right] -\frac{1-2k}{\left( 1-2k\right)
^{\frac{1}{2} } -\left( k-1\right) } \sinh \left[ f\left( x\right)
-f\left( a\right) \right] }{\frac{\left( 1-2k\right) ^{\frac{1}{2} }
}{\left( 1-2k\right) ^{\frac{1}{2} } -\left( k-1\right) } \cosh \left[
f\left( x\right) -f\left( a\right) \right] -\frac{1}{\left( 1-2k\right)
^{\frac{1}{2} } -\left( k-1\right) } \sinh \left[ f\left( x\right)
-f\left( a\right) \right] }\label{eq21}
\end{equation}
From formula{\nobreakspace}(\ref{eq21}) we conclude that for elastic scattering, 
i.e. when 
$k=\frac{1}{2} $, 
$A\left( x\right) =0$, and for 
$k\neq \frac{1}{2} $, 
$A\left( x\right) \neq 0$.

In the monograph{\nobreakspace}\cite{4} we introduced the relaxation time 
$\tau $ for quantum heat transport
\begin{equation}
\tau =\frac{\hbar }{mv^{2} }.\label{eq22}
\end{equation}
In formula{\nobreakspace}(\ref{eq22}) $m$ denotes the mass of heat carriers 
electrons and 
$v=\alpha c$, where 
$\alpha $
is the fine structure constant for electromagnetic interactions. 
As was shown in monograph~\cite{4}, 
$\tau $
 is also the lifetime for positron-electron pairs in vacuum.
 
When the duration time of the laser pulse is shorter than 
$\tau $, then to describe the transport phenomena we must use the hyperbolic 
transport equation. Recently the structure of water was investigated 
with the attosecond 
$\left( 10^{-18} s\right) $
 resolution{\nobreakspace}\cite{5}. Considering that 
$\tau \approx 10^{-17} $~s we argue that to study performed in \cite{5} open the new field for 
investigation of laser pulse with matter. In order to apply the 
equations{\nobreakspace}(\ref{eq9}) to attosecond laser induced phenomena we 
must know the cross section 
$\sigma \left( x\right) $. Considering formulae{\nobreakspace}(\ref{eq9}) and (\ref{eq22}) we obtain
\begin{equation}
\sigma \left( x\right) =\frac{mv}{\hbar } =\frac{me^{2} }{\hbar ^{2} }\label{eq23}
\end{equation}
and it occurs 
$\sigma \left( x\right) $
 is the Thomson cross section for electron-electron scattering.
 
With formula{\nobreakspace}(\ref{eq23}) the solution of Cauchy problem has the 
form:
\begin{eqnarray}
T_{+} \left( x\right) &=&\frac{2T_{0} e^{-\left( 1-2k\right)
^{\frac{1}{2} } \frac{me^{2} }{\hbar ^{2} } a} }{\left[ 1+\beta
e^{-2\left( 1-2k\right) ^{\frac{1}{2} } \frac{me^{2} }{\hbar ^{2} } a}
\right] } \times   \\ \label{eq24}
&&\mbox{}\left[\frac{\left( 1-2k\right) ^{\frac{1}{2} } \cosh \left[ \left(
1-2k\right) ^{\frac{1}{2} } \frac{me^{2} }{\hbar ^{2} }( x-a)
\right]}{\left( 1-2k\right)
^{\frac{1}{2} } -\left( k-1\right) } \right.\nonumber\\ 
 &&\mbox{}+\left.\frac{\left( k-1\right) \sinh \left[ \left( 1-2k\right) ^{\frac{1}{2} }
\frac{me^{2} }{\hbar ^{2} }( x-a) \right] }{\left( 1-2k\right)
^{\frac{1}{2} } -\left( k-1\right) }\right] ,\nonumber\\ 
T_{-} \left( x\right) &=&\frac{2T_{0} e^{-\frac{\left( 1-2k\right)
^{\frac{1}{2} } me^{2} a}{\hbar ^{2} } } \left[ \left( k-1\right) -\left(
1-2k\right) ^{\frac{1}{2} } \right] }{\left( 1+\beta e^{-2\left(
1-2k\right) ^{\frac{1}{2} } \frac{me^{2} }{\hbar ^{2} } a} \right) k}
\times \nonumber\\ 
&&\mbox{} \sinh \left[ \left( 1-2k\right) ^{\frac{1}{2} } \frac{me^{2}
}{\hbar ^{2} } \left( x-a\right) \right] .\nonumber
\end{eqnarray}
\section{Conclusions}
In this paper the one dimensional Weyl type thermal equation 
was developed and solved. It was shown that depending on the 
dynamics of the heat carriers scattering the damped or oscillated 
temperature field can be generated. When the laser pulse generates 
relativistic electrons the cross section for the generation of 
electron-positron pairs is equal to the Thomson cross section.


\newpage
\begin{thebibliography}{9}
\bibitem{1}E. L. Saldin et al., http://lanl.arxiv.org/physics/0403067 .
\bibitem{2}W. Greiner, \textit{Relativistic Quantum Mechanics}, Springer Verlag, 
Berlin, 1990.
\bibitem{3}J. D. Bjorken and S. D. Drell, \textit{Relativistic Quantum Mechanics}, 
McGraw Hill, New York, 1964.
\bibitem{4}M. Kozlowski, J. Marciak-Kozlowska, \textit{From Quarks to 
Bulk Matter}, Hadronic Press, USA, 2001.
\bibitem{5}P. Abbamonte et al., \textit{Phys. Rev. Letters}, \textbf{92}, 
(2004), p. 237401-1
\end{thebibliography}
\end{document}